\documentclass[preprint,10pt,numbers]{sigplanconf}

\usepackage{amsmath}

\usepackage{graphicx}
\usepackage{epsfig}
\usepackage{courier}
\usepackage{threeparttable} % Soporta la inserción de notas al pie de la tabla.

\begin{document}

\setlength{\pdfpageheight}{\paperheight}
\setlength{\pdfpagewidth}{\paperwidth}

\conferenceinfo{CONF 'yy}{Month d--d, 20yy, City, ST, Country}
\copyrightyear{20yy}
\copyrightdata{978-1-nnnn-nnnn-n/yy/mm}
\copyrightdoi{nnnnnnn.nnnnnnn}

% Uncomment the publication rights you want to use.
%\publicationrights{transferred}
%\publicationrights{licensed}     % this is the default
%\publicationrights{author-pays}

\titlebanner{banner above paper title}        % These are ignored unless
\preprintfooter{short description of paper}   % 'preprint' option specified.

\title{HADAS Green Assistant: designing energy-efficient applications}
%\subtitle{Subtitle Text, if any}

\authorinfo{Nadia G\'amez\and M\'onica Pinto\and Lidia Fuentes}
           {CAOSD Group, University of M\'alaga, M\'alaga, SPAIN}
           {\{nadia,pinto,lff\}@lcc.uma.es}
%\authorinfo{Name2\and Name3}
%           {Affiliation2/3}
%           {Email2/3}

\maketitle

\begin{abstract}
%The information and communication technologies (ICTs) are shaping the future, but software designers are still not aware about the impact of the energy consumed by these systems in the Earth’s greenhouse effect.
The number of works addressing the role of energy efficiency in the software development has been increasing recently. But, designers and programmers still complain about the lack of tools that help them to make energy-efficiency decisions. Some works show that energy-aware design decisions tend to have a larger impact in the power consumed by applications, than code optimizations. In this paper we present the HADAS green assistant, which helps developers to identify the energy-consuming concerns of their applications (i.e., points in the application that consume more energy, like storing or transferring data), and also to model, analyse and reason about different architectural solutions for each of these concerns.
%So, the developer can choose among different alternatives for a particular energy-consuming concern, and then, analyse and reason about the energy impact of each design decision.
This tool models the variability of more or less green architectural practices and the dependencies between different energy-consuming concerns using variabilty models.
%It also models the architectural dependencies between different energy-consuming concerns, meaning that the computation of the expected energy consumption of one of those concerns, also considers other concerns needed to apply a specific architectural pattern.
Finally, this tool will automatically generate the architectural configuration derived from the selections made by the developer from an energy consumption point of view.
\end{abstract}

\category{D.2.11}{SOFTWARE ENGINEERING}{Software Architectures}

\keywords Energy-efficiency, Software Architectures

\section{Introduction}

Energy-aware software development (or Green Computing \cite{Li2011}) is a growing trend in computing. Indeed, the increasing number of papers addressing software sustainability in last years clearly indicates that today software developer community is starting to pay more and more attention to the energy-efficiency concerns.

However, recent empirical studies \cite{Pinto2014,Pang2015,Manotas2016,Chitchyan2016} show that software developers do not have enough knowledge about how to reduce the energy consumption of their software solutions. The majority of developers are not aware about how much energy their application will consume and so, they rarely address energy efficiency \cite{Pinto2014,Pang2015}. Even practitioners that appear to have experience with green software engineering have significant misconceptions about how to reduce energy consumption \cite{Manotas2016}.
%Indeed the main causes that they identify as the most influential on energy consumption are not always correct \cite{Pinto2014,Pang2015}.
Also, software developers are unsure about the patterns and anti-patterns associated to energy-efficiency \cite{Manotas2016}. These studies also evidence the lack of tool support of green computing, not only at the code level, but also at higher abstraction levels –--i.e., requirements and software architectures levels \cite{Chitchyan2016}. The main conclusion of these studies is that software developers need more precise evidence about how to tackle the energy efficiency problem and some tool support that help them to effectively address it \cite{Pang2015,Manotas2016}.

There are plenty of experimental approaches that try to identify what parts of an application influence more in the total energy footprint of an application --i.e., to identify the \emph{energy hotspots} \cite{Noureddine2015-2}. An important part of these works proposes to minimize energy consumption by focusing on code level optimizations. They report the energy consumption of different implementations. For example, of data collections in Java \cite{Hasan2016}, or system calls in Android applications \cite{Li2014}. There are however other works that demonstrate that changes in architecture design tend to have a larger impact in energy consumption \cite{grosskop2013}. However, analysing the expected energy consumption of so many alternative architectural solutions is not a trivial task. Developers would need tool support that helps them to measure, analyse and reason about alternative architectural solutions to energy hotspots ---i.e., the set of components that implement a given energy hotspot (hereinafter, \emph{energy-consuming concerns}).

One of the benefits of addressing the energy efficiency at architectural level is to provide software developers with the necessary means to analyse the energy consumption of different alternative solutions, before implementing them. Energy absolute values are not needed, because what is important for developers is to be able to compare the energy consumed by different architectural alternatives \cite{JagroepWBPLBV16}. There is no doubt that the green computing community has made many steps forward in the development of green software architectures. Some relevant examples are the catalogs of energy-aware design patterns \cite{Noureddine2015-1} and architectural tactics \cite{Procaccianti2014}, as well as new architecture description languages that incorporate an energy profile and analysis support \cite{Stier2015, Ouni2012}.

However, we argue that there is still not enough tool support that helps developers to clearly identify the energy-consuming concerns in their applications, and moreover to choose and generate the most appropriate architectural solutions from an energy consumption point of view. On the one hand, recent studies although complementary, are disconnected, so their results cannot be easily applied to the development of green applications in an integrated way \cite{Manotas2014}. On the other hand, these studies do not always consider that some energy-consuming concerns (e.g., to store data locally or remotely) have strong dependencies with others (e.g., to store data remotely will depend on the communication concern, and the latter one, on the security concern) \cite{DeMaio2016}. Moreover, usually the results of these studies are not easily accessible for practitioners, which do not know how to apply and reuse this knowledge in their applications \cite{Pinto2014,Pang2015,Manotas2016,Chitchyan2016}. In order to cope with these limitations we consider that software developers need some kind of 'assistant' that guides them through all the steps required to identify, model, analyse and reason about different architectural solutions to energy-consuming concerns.

In this paper we present the \emph{HADAS green assistant}, that aims
to help developers to generate the most energy-efficient
architectural configurations that fulfil the application
requirements. This assistant will suggest a set of energy-consuming
concerns (i.e., points in the application that consume more energy,
like encrypting or transferring data) and for each of them it will
show the list of possible architectural solutions, along with an
energy function. Each of the provided solutions were previously
modeled and their energy consumption calculated or predicted, before
storing them in the repository (for predictions we use Palladio
Power Consumption Analyzer \cite{Stier2015}). So, HADAS drastically
reduces the effort of analysing the energy consumed by different
architectural solutions, which in other works has to be performed
from the scratch. HADAS internally models the variability of the
architectural solutions for the energy consuming concerns using
variability models, concretely the Common Variability Language (CVL)
\cite{HaugenWC13}  (e.g., cache memory can be modeled as a optional
feature). It also models the architectural dependencies between
different energy-consuming concerns, meaning that the computation of
the expected energy consumption of one of those concerns, also
considers other concerns needed to apply a specific architectural
pattern (this is not always considered in other works).

To sum up, with the HADAS green assistant, the developer can choose among different alternatives for a particular energy-consuming concern (e.g. storing information, communication or compression) and will be able to analyse and reason about the energy impact of each design decision. Finally, this tool will automatically generate the architectural configuration derived from the selections made by the developer from an energy consumption point of view.

%MONICA: Otra forma sería poner las contribuciones tipo lista. Se repetiría con lo anterior, por lo que hay que elegir una de las dos formas:

%The main contributions of the HADAS green assistant are repository are: (1) Puede ayudar a mejorar el conocimiento que los des. del software tienen (cuales son realmente los energy-consuming concerns, como afectan sus decisiones, qué dependencias existen y que hay que tener en cuenta para realizar el análisis. Ej: el almacenamiento de datos de forma comprimida puede ser más costoso, pero si el coste es menor que el envío a través de la red, merece la pena asumir ese coste. No se pueden analizar las decisiones de forma independiente, sino de forma conjunta haciendo uso de las dependencias. No necesita saberlas, nosotros se lo proporcionamos)- eliminar misconceptions; (2) ... and (3) ...

After this introduction and the related work described in Section 2,
the requirements to implement the HADAS green assistant are detailed
in Section 3. Then, in Section 4 the HADAS green repository and the
HADAS green assistant are described from two different points of
view. One is the use of the green assistant by the software
developers of green applications. The other one is the technical
details regarding their implementation. Our approach is evaluated
and the results discussed in Section 5. Finally, the conclusions and
on-going work are described in Section 6.

\section{Related Work}
In order to better motivate our work, we have reviewed papers focused on: (1) experimental studies at the code level (CL); (2) proposals at requirement (RL), architecture (AL), and design levels (DL); and (3) studies about energy consumption awareness of software developers. Due to the large number of existing works, and the rapid changes in this area, we will narrow our study to those papers published in last editions of relevant (energy-specific) software engineering conferences and journals. We consider these are representative papers of current research in this area.

Table 1 summarizes the different papers we have considered in this
section. For each of them we indicate the \emph{level} at which the
paper focuses (second column), the \emph{type} of paper (third
column), whether \emph{dependencies} are considered or not (fourth
column), the main \emph{output} (fifth column) and the
\emph{knowledge} that is derived from this work (sixth column).

\begin{table*}
\centering
\caption{Related Work: Energy-aware approaches}
\label{tab:relatedWork}
 \begin{threeparttable}
\begin{tabular}[c]{|p{2.3cm}|p{.7cm}|c|c|p{5.4cm}|p{5.4cm}|}
    \hline
    \textbf{Proposal} & \textbf{Level} & \textbf{Type} & \textbf{Dep.} & \textbf{Output} & \textbf{Knowledge} \\ \hline
    1. Hasan \cite{Hasan2016} & CL & Exp. & No & Energy profiles of operations on Java Collections. & It claims that a per-method analysis of energy consumption must be made. \\ \hline
    2. Li \cite{Li2016} & CL & Exp. & No &  Multiple HTTP requests are bundled to reduce energy consumption. & HTTP requests are one of the most energy consuming operations. \\ \hline
    3. Chen \cite{Chen2015} & CL & Exp. & No & Performance and energy consumption profiles for cloud applications. & Tool support is needed; also using realistic application workloads. \\ \hline
    4. Li \cite{Li2014} & CL & Exp. & No & Quantitative information about the energy consumption of Android apps (405 apps) & More than 60\% of energy consumed in idle states; the network is the most energy consuming component; developers should focus on code optimization.\\ \hline
    5. Procaccianti \cite{Procaccianti2016} & CL & Exp. & No & Report on two green software practices: use of efficient queries and put applications to sleep.
    & Software design and implementation choices significantly affect energy efficiency. The effectiveness of best practices for reducing energy consumption needs to be precisely quantified.  \\ \hline
    6. Manotas \cite{Manotas2014} & CL & Fram. & No & Define the SEED framework for the automatic optimization of energy usage of applications by making code level changes. & Support is needed to integrate the insights gained by existing experimental studies to help identifying the more energy-efficient alternatives.
     \\ \hline
    7. Noureddine \cite{Noureddine2015-1} & DL CL & Exp. & No & Empirical evaluation of 21 design patterns. Compiler transformations to detect\&transform patterns during compilation for better energy efficiency with no impact on coding practices. & The energy consumption of design patterns highly depend on the running environment; several studies identified both patterns and anti-patterns regarding energy consumption.\\ \hline
    8. Procaccianti \cite{Procaccianti2014} & AL & Mod. & Yes & Identify energy efficiency as a quality attribute and define green architectural tactics for cloud applications. Identify relationships between different architectural tactics. & Energy efficiency has to be addressed from a software architecture perspective. Software architects need reusable tactics for considering energy efficiency in their application designs. \\ \hline
    9. PCM \cite{Stier2015} & AL & Mod. & Yes & Architecture Description Language and tool set with support for the specification, calculation and analysis of energy consumption. & At the architectural level the energy consumption can be estimated based on resource consumption (CPU, HDD, etc.) and usage models.
     \\ \hline
    10. AADL \cite{Ouni2012} & AL & Mod. & Yes & Plug-in integrated with the AADL tool to support the specification and analysis of energy consumption. & It focuses in the energy overhead of inter-process communication, an important service of embedded systems. \\ \hline
    11. Pinto \cite{Pinto2014} & CL & EmpSt & - & Qualitative study exploring the interest and knowledge of software developers about energy consumption & Lack of tools; many misconceptions and panaceas. Major causes for energy consumption problems are identified.
      \\ \hline
    12. Chitchyan \cite{Chitchyan2016} & RL & EmpSt. & - & Qualitative study exploring requirements engineering practitioners’ behaviour towards sustainability (including energy consumption). & Lack of methodological support; lack of management support; requirements trade-off and risks, ... \\ \hline
    13. Manotas \cite{Manotas2016} & RL AL CL & EmpSt. & - & Qualitative study exploring the knowledge of practicioners interested on energy consumption from different perspectives (requirements, design and construction). & Green software practitioners care and think about energy; however, they are not as successful as expected because they lack necessary information and tool support. \\ \hline
    14. Pang \cite{Pang2015} & CL & EmpSt. & - & Qualitative study exploring the knowledge of practicioners about energy consumption. & Programmers rarely address energy. There are important misconceptions about software energy consumption.
 \\ \hline
\end{tabular}
\begin{tablenotes}
    \item CL - Code Level, AL - Architecture Level, DL - Design Level, RL - Requirements Level
    \item Exp. - Experimental Work, Fram. - Framework, Mod. - Modelling Work, EmptSt. - Empirical Study (Questionnaires)
\end{tablenotes}
\end{threeparttable}
\end{table*}

Firstly, in rows 1 to 7 we can observe that a large number of papers present experimental studies performed at the code level. A common goal to all of them is the definition of energy profiles for different energy-consuming concerns. \emph{These experimental studies usually focus on one particular energy-consuming concern (e.g. communication or data storage) without considering the dependencies among them}.
%(e.g. there are dependencies between communication, compression and data storage because the energy consumption may vary when a file is stored locally or remotely or depending on the compression algorithm).
It is important to highlight that a \emph{few of these proposals provide some support to integrate/reuse the knowledge obtained from experimental studies}. For instance, the work in \cite{Procaccianti2016} (row 5) defines a wiki with a template to integrate all the results from different experimental studies and the work in \cite{Manotas2016} (row 6) defines a framework to integrate different energy-efficient alternatives. \emph{But none of them, model explicitly the variability of energy consuming concerns, loosing the opportunity to automatically generate and manage green product configurations}.

There are also an increasing number of proposals that focus at higher levels of the software development, such as design (row 7) or architecture (rows 8 to 10). Focusing on high level proposals, we can distinguish between experimental works (row 7) and modelling works (rows 8 to 10). Some of these modelling works are architecture description languages that provide support for analyzing energy consumption (rows 9 and 10). Most of the works at architectual level consider the relationships between different energy concerns, although in different ways. For instance, in \cite{Procaccianti2014} authors define the joint use of different architectural tactics. Also, the works in \cite{Stier2015} and \cite{Ouni2012} provide support to specify the relationships between components modeling different energy concerns. These relationships can then be used during the analysis phase to see how a energy concern (e.g. communication) can influence in other energy concerns (e.g. compression or data storage). \emph{But, the identification and specification of dependencies between energy concerns has to be done manually by the software engineer}.

%MONICA: No es que manejen las dependencias de la misma forma que nosotros lo hacemos, pero tampoco creo que fuera correcto decir que ninguna propuesta tiene en cuenta ninguna dependencia. En este caso los revisores podrían hacer ver que eso no es correcto. No se exactamente si expresarlo así o de otra forma, pero creo que habría que reconocer que algunas propuestas a nivel de arquitecturas ya identifican, aunque de forma distinta a como nosotros lo hacemos, esas relaciones entre distintos elementos de la arquitectura que afectan al consumo de energia

Finally, the number of empirical studies at different stages of the software life cycle (requirements, design, implementation) and with different groups of software developers (e.g.  worried (or not) about the energy consumed by their applications) are considerably increasing in last years (see rows 11 to 14). As indicated in the introduction, the results of all these empirical studies are the same: \emph{there is a lack of methodological and tool support and software developers have still many misconceptions about how to reduce the energy consumption of their applications}.

\section{Green Assistant Requirements}
The implementation of a green assistance implies addressing the following requirements:

\begin{itemize}
%MONICA: Se me hacía difícil de entender, porque había muchos "Additionally" y "Also". Lo he cambiado ligeramente
\item \textbf{R1: Identify and model the energy hotspots}. Since developers do not have yet enough knowledge about what concerns could impact more in the power consumption, the green assistant should support them in this task. Although recent works propose different architectural tactics to implement concrete energy hotspots, none of them explicitly model the architectural variability of the energy consuming concerns. Additionally, the energy consuming concerns associated with every energy hotspot could depend on other energy consuming concerns. But these kinds of dependencies often go unnoticed by the developer, and so they do not include them in the energy analysis. Then, the green assistant should automatically include those energy consuming concerns that depend on the ones already selected by the developer. The variability models and dependencies should also be part of the models stored in the green repository.
\item \textbf{R2: Design the architecture of every valid configuration and resource consumption of each component}. Up to the moment, if a developer wants to know the resource consumption of a concrete architectural solution that fits an energy hotspot, they need to manually specify the architecture and calculate the resources needed by each component (for example, with Palladio). Considering that for a given energy hotspot there could be many solutions, designers would not be interested in doing this for every alternative. But, without this information it is not possible to guide the developer in selecting the most energy efficient solution. So, the great challenge here is to provide the developer with a pre-defined architecture for each of the variants of every energy consuming concern, and an estimation of the consumed resources. The benefit is twofold: (i) the developer knows in advance the resources needed by the energy consuming concerns of their application, simply clicking a button; (ii) the architectural design of the selected solution can be reused, being part of the final architecture of the application. These models should also be part of the green repository.
\item \textbf{R3: Calculate the energy function of each architectural configuration}. We have already seen that the energy consumed by an application usually depends on input parameters, but the challenge is who is going to define the energy function for a concrete architectural configuration. The only way of making this is manually, so ideally the green assistant should already provide this information. Since an application usually has to include many energy consuming concerns, having the energy function of each of them previously calculated will help designers to see the power consumption of the final application, and make some corrections in their decisions if it is necessary. Finally, this information should enrich the models stored in the green repository.
\item \textbf{R4: Implement the user interface of the green assistant tool}. The approach presented in this paper is not viable without a user interface. This user interface should show: (i) the list of energy consuming concerns associated with the energy hotspots; (ii) the options available and a mechanism to enable and disable other options (i.e., other energy consuming concerns) according to the dependencies identified in R1; (iii) a energy efficiency analysis with graphics showing the energy consumption in function of some input parameters;(iv) an option to generate the architectural configuration corresponding to the selections made by the developer from the energy consumption point of view.

\end{itemize}

\section{The HADAS Green Assistant}
In this section we will present the HADAS Green Assistant tool, focusing especially on describing the HADAS Green Repository. They are described both from the point of view of the software developers who want to use it and from a technical perspective. %So, we will show how to use it and we will present the complexity of implementing such a tool.
Suppose that Alice is a
software developer that wants to develop an energy-efficient
application. We will use a Media Store application, previously
implemented and defined in Palladio, to illustrate our proposal and
the advantages of using our tool for choosing energy efficient
architectures adapted to the requirements.

\begin{figure}
\includegraphics[width=\columnwidth]{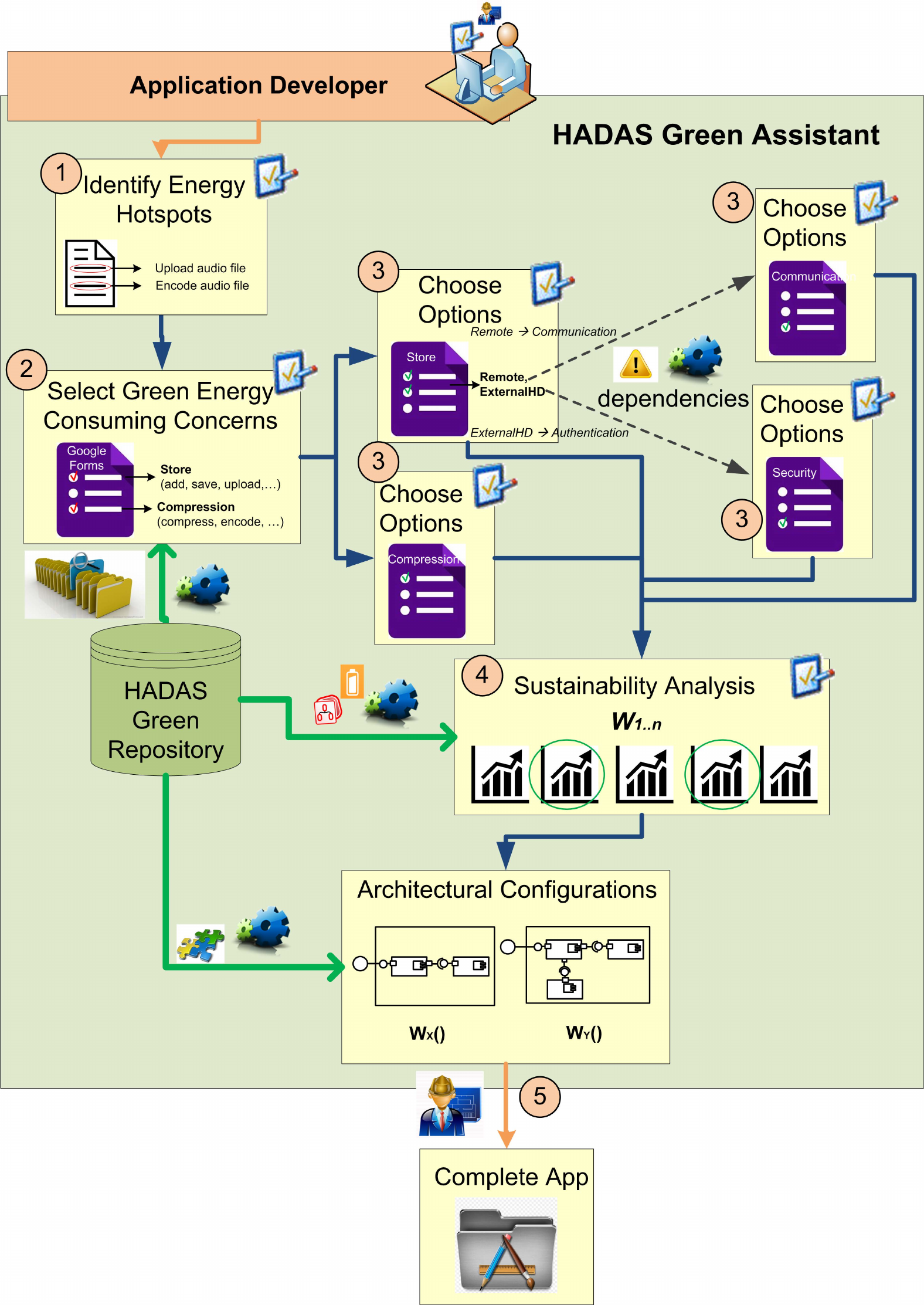}
\caption{An Schema of the HADAS Green Assistant tool}
\label{fig:assistant}
\end{figure}

If Alice wants to use the HADAS Green Assistant, she must follow the
steps shown in Figure \ref{fig:assistant}. \emph{Note that we will
describe these steps in italic letter in order to differenciate
between the user point of view and the implementation details}.

\subsection{Energy Consuming Concerns}

\emph{The \underline{first step (Figure \ref{fig:assistant}, label 1)} is to identify
energy hotspots in the application requirements. Alice is going to
develop the Media Store application, so she needs to store audio
files in a server and/or to encode these files, among other
functionality}.

\begin{figure}
\includegraphics[width=\columnwidth]{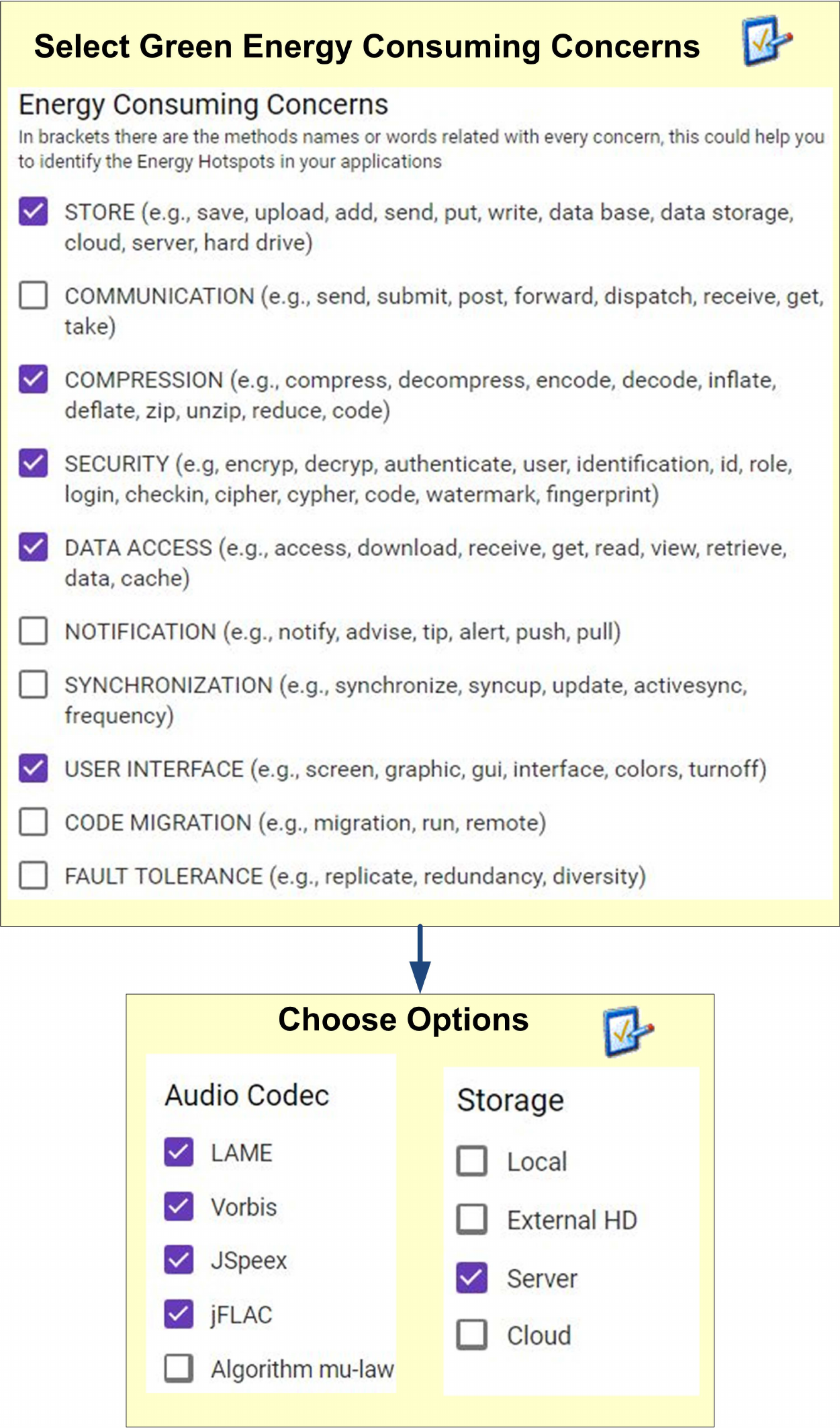}
\caption{HADAS Green Assistant interface: some Forms}
\label{fig:forms}
\end{figure}

The HADAS Green Assistant has previously identified these kind of
concerns as energy consuming. For instance, \textsf{Store} (e.g., upload an
audio file) and \textsf{Compression} (e.g., encode an audio file). We have
explored nowadays applications in order to identify the main energy
hotspots that are repeated among many of these applications. So far
our repository have a list of 10 energy consuming concerns, as can
be seen in Figures \ref{fig:forms} and \ref{fig:repositoryCVL}:
\textsf{Store}, \textsf{Communication}, \textsf{Compression}, \textsf{Security}, \textsf{Data Access},
\textsf{Notification}, \textsf{Synchronization}, \textsf{User Interface}, \textsf{Code Migration} and
\textsf{Fault Tolerance}. Of course, this list will be augmented when we have
new evidence about other energy hotspots.

\emph{The \underline{second step (Figure \ref{fig:assistant}, label 2)} for Alice is
to select the energy-consuming concerns in the HADAS respository}.

The user interface of the HADAS assistant is implemented as a Google
Form, that fits very well what we need. This Google Form (see Figure
 \ref{fig:forms}) has a list with all the concerns and although they are pretty intuitive, in
the form we also show some keywords that could be associated
with each of these concerns. For example, the \textsf{Store} concern could be
used when in the application requirements the developer finds words
like \textsf{save}, \textsf{upload}, \textsf{add}, \textsf{send}, \textsf{put}, \textsf{write}, among others. So,
applications developers can identify easily which concern correspond
to their application hotspots.

\emph{In the requirements of the Media Store, Alice can see words like
storage, upload (corresponding with the \textsf{Store} concern); download, cache
(corresponding with the \textsf{Data Access} concern); users, login, encrypted
(corresponding with the \textsf{Security} concern); encode, compressed
(corresponding with the \textsf{Compression} concern); GUI, interface (corresponding
with the \textsf{User Interface} concern). So, now Alice knows that these
concerns may strongly affect the final energy expenditure of her
application and decides to perform the sustainability analysis offered by the HADAS assistant to compare different architectural configurations and choose the one that better fits her needs. Then, she should select all the energy-consuming
concerns provided by HADAS that she needs to develop the
application. In Figure \ref{fig:forms} it can be seen that she has
selected the \textsf{Store}, \textsf{Compression}, \textsf{Security}, \textsf{Data Access} and \textsf{User
Interface} concerns}.

There exists many variability in the way to design and implement
these concerns (e.g., the data could be stored locally or remotely,
there are many encryption algorithms, or different codecs to
compress audio or video files). Additionally, these concerns are not
independent from each other. For instance, there are several
concerns related with \textsf{Communication}, such as \textsf{Data
Access}, \textsf{Store}, \textsf{Notification},
\textsf{Synchronization} or \textsf{Code Migration}. This means that
there are dependencies between them and, therefore, the energy
consumption cannot be analyzed isolatelly for every concern.
Instead, a whole architecture should be analyzed, where these
dependencies are explicitly modeled and taken into account. In the
next subsection we will see how this variability and the dependences
are modeled in our approach.

\subsection{Variability Model and Dependencies }

\emph{The \underline{third step for Alice (Figure \ref{fig:assistant}, label 3)} is
to choose the options for every energy-consuming concern selected in the second step.} The HADAS assistant will show the application developer a separated google
form for every selected concern and in our case \emph{Alice has to select
the alternatives she wants to explore to analyze later the energy
efficiency.} For instance, the store could be done locally and/or
remotely in an external hard drive, in a server or in the cloud.
Depending on this selection, alternatives for other energy consuming
concerns could also be suggested by the assistant. \emph{The Media Store
application will store the audio files in a \textsf{Server}, so Alice has selected this option for the \textsf{Store}
concern (as can be seen
in Figure \ref{fig:forms})}. This selection implies that the audio files have to be uploaded
to a remote server. The \textsf{Communication} concern, which is the
responsible for sending the file, was previously identified by HADAS
as another energy consuming concern and it is included in the
repository. Thus, communication is required to upload the files and must be included in the analysis in order to know how much energy will be
consumed by the whole architecture. However, \emph{Alice was not aware of the dependency between the remote storage in a server and communication and, thus, she did not explicitly selected the \textsf{Communication} concern during the first step (as can be seen in Figure \ref{fig:forms})}. However, this is not a problem when using HADAS because, in order to make easier the developer job, the HADAS assistant will automatically show the options available for the \textsf{Communication}
concern. This is possible because the HADAS repository contains information about the dependencies between different energy concerns.

We have modeled the energy consuming concerns variability and the
dependencies between them using the Common Variability Language
(CVL). CVL allows modelling the variability separately from a base
model (i.e. architectural model), but both the variability and the base models can be connected
and can be managed using the same tool. In particular, using the CVL
tools we specify the Variability Model (called VSpec tree) and the
binding between this and the Base Model. With a VSpec tree we can
specify the common features that must be part of the architectural
solution for a given energy consuming concern, and also their
variants (e.g., to store data locally or remotely).

\begin{figure*}
\includegraphics[width=\textwidth]{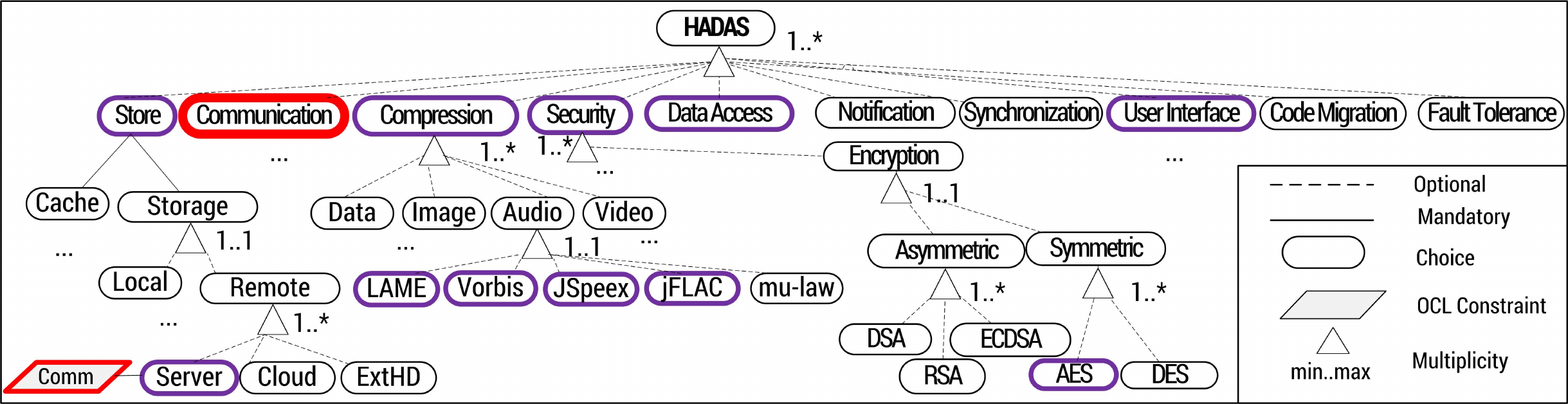}
\caption{The variability model of the HADAS Green Repository in CVL}
\label{fig:repositoryCVL}
\end{figure*}

Some of the features of this variability model can be seen in Figure
\ref{fig:repositoryCVL}. We have depicted part of the variability
for some concerns related to the Media Store case study. The rest
were not included for the sake of simplicity and also because it is
out of the scope of this paper. Specially, Figure
\ref{fig:repositoryCVL} shows part of the variability for the
\textsf{Store} concern, the variability of the audio codecs of the
\textsf{Compression} concern and also the \textsf{Encryption}
variants of the \textsf{Security} concern. We have put in bold the
lines of the selected features that correspond to Alice's checkbox
selections in the forms shown in Figure \ref{fig:forms}.
\emph{Concretely, Alice has selected to store the files in a
\textsf{Server} and to explore four different codecs or algorithms
for the audio compression: \textsf{LAME}, \textsf{Vorbis},
\textsf{jFLAC} and \textsf{JSpeex}}.

Note that \textsf{Communication} is also selected in the variability model of \ref{fig:repositoryCVL} (marked in a
bolder red line). This is because Alice selected the \textsf{Server} feature of the \textsf{Store
concern}, and there is a cross-tree constraint between this and the \textsf{Communication} concern that implements the dependency between both concerns. This constraint is called \textsf{Comm} in the figure and it is marked in red. It formally defines the mutual dependency as: \texttt{Server implies Communication}.

With CVL it is possible to generate automatically the Resolution
Models after selecting (True or False) a set of choices in the
variability model. These selections in the variability model are
obtained from the selections in the google forms. A Resolution Model
represents a set of alternatives for every selected concern that the
developer wants to explore during the analysis of the power
consumption. Every alternative corresponds with a concrete
architecture of the base model. We will detail how we define this
architecture in next subsection. The benefit of using the HADAS
assistant is that the specifications of the variability model and
the generation of the resolution model, as well as the specification
of the alternative software architectures, are steps completely
transparent to Alice, who only has to deal with the google forms.

\subsection{Architecture and Energy Consumption Simulations}

\emph{Once Alice has selected the energy consuming concerns and the
variants for the ones she wants to explore, she has to make an
energy-efficiency analysis of the different alternatives (\underline{Figure
\ref{fig:assistant}, label 4) as the fourth step}. HADAS will help her to be aware of
the energy implications of their decisions}.

The energy consumption of each concern variant is expressed by means
of energy functions parameterized by variable parameters (e.g., the
audio file size). HADAS does not pretend to provide the exact
consumption in Watts, because what Alice needs to know is which
alternatives are more energy consuming than others, and which ones
can be considered more green. \emph{This is an iterative process,
since Alice can select different options for several concerns and
analyse the impact of each of her decisions in the energy
expenditure of the application.} So, the HADAS Green Assistant will
show the energy consumption for every concern but also for the whole
architecture, since as mentioned before the concerns are not
independent from each other. Then, as we describe in the next
section,  we need to be able to simulate the energy expenditure and
of all the possible architectural alternatives of all the concerns.
In this way the HADAS Green Assistant will provide the energy
functions for every configuration generated from the variability
model described in the previous subsection.

\begin{figure}
\includegraphics[width=\columnwidth]{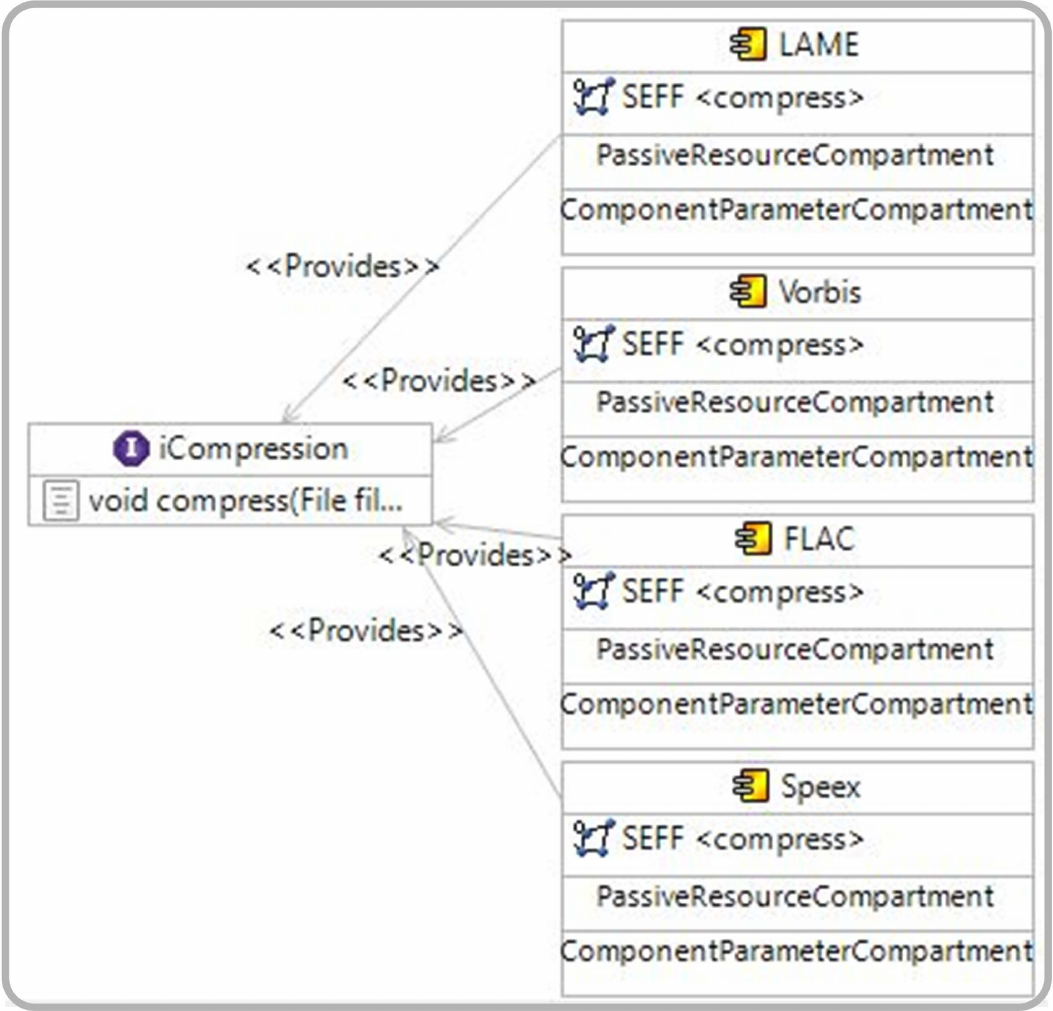}
\caption{PCM Repository Diagram for the Compression Concern}
\label{fig:compressionpcm}
\end{figure}

The Base Model of CVL must be a MOF compliant model of the software
architecture, which in our case is the set of components and
connections that specify a concrete concern variant. Since we need
to include the expected energy consumption of each variant, we have
to use an architectural model or language that provides this kind of
information. We have chosen to model the architecture using the
Palladio Component Model (PCM) due to the powerful toolset that
provides (Palladio) to analyse the resource consumption at
architectural level (including the energy). The metamodel of PCM can
be implemented in MOF, so it can perfectly be used jointly with CVL.
Therefore, in order to automatically provide the energy consumption
functions for the configurations we connect the CVL variability
model (VSpec Tree) with the respective architectural base model
specified in PCM. Figure \ref{fig:compressionpcm} depicts some of
the components that provides the \textsf{Compression} interface to
compress audio files of different formats using different algorithms
or codecs.

Then, using the Palladio Power Consumption Analyzer we simulate the
different configurations to obtain the power consumption. Thus, the
application developer will automatically know how the energy
consumption varies when they select different alternatives for the
selected concerns. The total number of alternatives when there are
several selected concerns could be really high, so reason about and
simulate the energy expenditure of all these possible alternatives by
hand for every application is not possible. HADAS helps to that by previously simulating the energy expenditure
of every possible configuration of our repository with Palladio.
This takes a lot of time, but with our approach it has to be performed only once. The
benefit for the developer is that they do not have to make any
resource consumption simulation by hand. The tool will show the
results so that the developer only has to analyse the different
results.

\emph{Furthermore, thanks that HADAS takes into account the
dependencies between different energy consuming concerns, Alice will
be able to make complex decisions with no so much effort. For
example, Alice can decide not to include the possibility to upload
compressed audio files when they are too small (less than 5 Mb). She
was able to make this decision, because the energy consumption
functions provided by HADAS take into account both the energy
expenditure to compress the file and the energy expenditure to send
the compressed file to the server.} This has been possible because
HADAS exploit the dependencies between the concerns. Then although
in the Media Store requirements Alice did not identified
communication as a energy hotspot, HADAS automatically suggested her
to consider also this concern.
%MONICA: Esta última parte se repite con una frase que he añadido en la sección 4.2. Habría que ver donde sería mejor añadirla porque realmente es en la 4.2 donde se habla de las dependencias. Aquí se podría hacer una referencia a ello pero sin hacer tanto hincapié.
It showed the different energy functions for the
upload functionality including also the compression (specified as another cross-tree constraint).
This is very important since, as will be detailed in next section, the
decisions taken could be different if we analyse the audio compression
without considering the communication and viceversa.

\emph{The \underline{last step Alice has to perform (Figure \ref{fig:assistant}, label 5)} is to get the architectural
configuration from the HADAS repository simply clicking a button}. For
that, HADAS will generate the architectural configuration including
the definitive options Alice has selected after the analysis in the step 4) using our implementation of the CVL engine. For instance, HADAS will include in the configuration (i.e. the resolved model in CVL) components for the
\textsf{Store}, \textsf{Compression}, \textsf{Data Access}, \textsf{Security} and \textsf{Communication} concerns corresponding to the
options chosen by her. \emph{In this case Alice, will have to
complete the application with the rest of functionality
considered not relevant for energy efficiency (e.g., audio file
edition)}.

\section{Evaluation}
In this section we are going to detail how we perform the
simulations that are needed to be able to inform,  through the HADAS
green assistant, about the power consumption of the alternative
architectures software developers want to explore. We perform the
simulation in advance to store in the HADAS green repository an
energy function that will provide how many watts every one of the
possible configuration of each concern are going to waste, at least
in relative terms -- i.e., which one of the different variants waste
less and which one more. We differentiate two main steps. A first
step where we perform the experiments to obtain the energy functions
for each energy-consuming concern. And a second step where we
integrate the information from the experiments into the Palladio
Consumption Analyzer in order to simulate the joint use of several
dependent concerns.

\subsection{Experimental studies}
For every energy-consuming concern we have carried out a set of experiments
measuring the energy waste with Joulmeter\footnote{https://www.microsoft.com/en-us/research/project/joulemeter-computational-energy-measurement-and-optimization/},
a Microsoft modeling tool to measure the energy usage of software
applications running on a computer.  This tool has been calibrated
using Watts'Up\footnote{https://www.wattsupmeters.com/secure/products.php?pn=0} to be able
to obtain the real power consumption of every hardware component
(e.g., CPU, HDD, Screen, ..). All the experiment has been conducted
in a Gateway DT30 Desktop PC with Intel Core 2 Quad Q9300, 2.50GHz
and 8GB of RAM under Windows 10, 64 bits. And all the concerns have
been implemented in Java.

\begin{figure}
\includegraphics[width=\columnwidth]{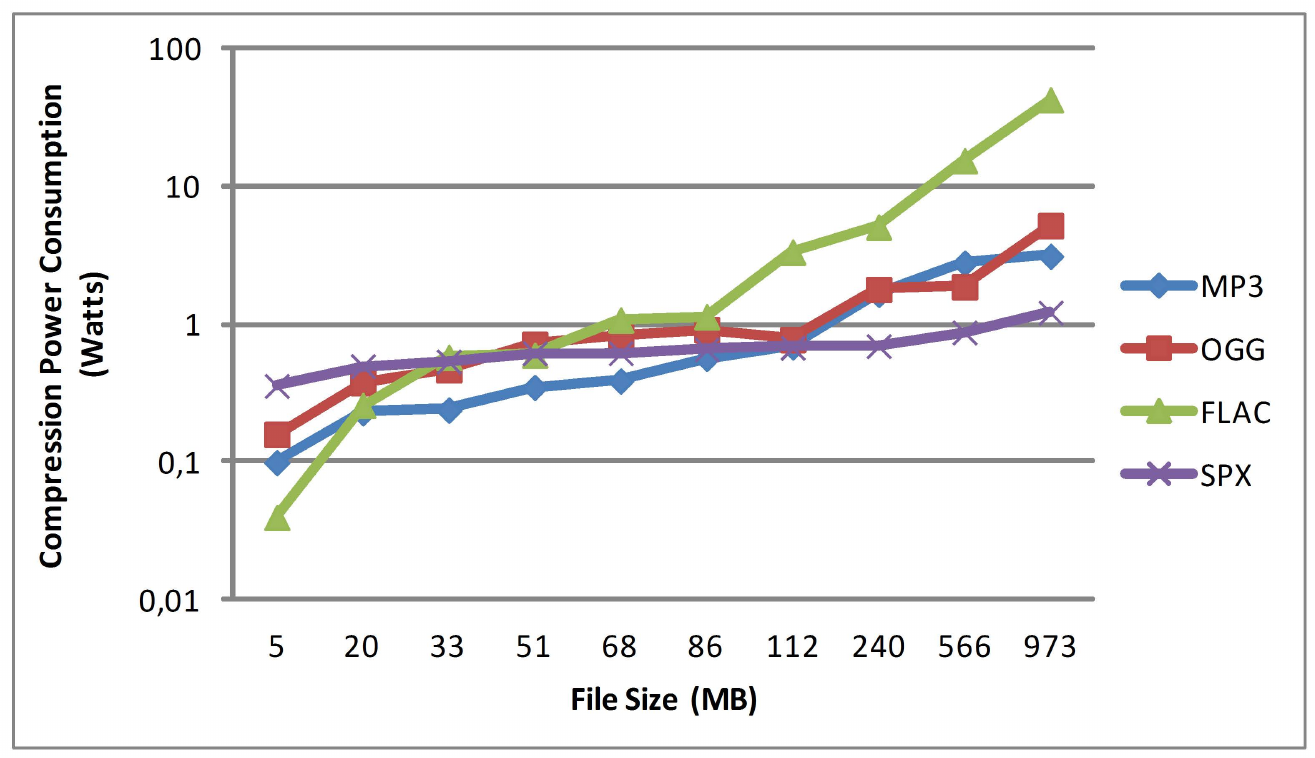}
\caption{Compression Power Consumption}
\label{fig:graphics_compression}
\end{figure}

Remember that Alice chose 4 different audio codecs to compress audio
files because she wanted to know which one(s) were more appropriate,
from a energy point of view, to be included in her Media Store
application. So, in order to let her know which one is more energy
efficient for her application, in the step 4 described in previous
section, the HADAS green assistant could show the graphic of Figure
\ref{fig:graphics_compression}. This graphic shows the power
consumption (in a logarithmic scale) to compress 9 WAV audio files
of different sizes (from 5Mb to almost 1GB) using the following
audio compression algorithms: Java LAME 3.99.3 to create MP3 audio
files using a bitrate of 128Mb, Vorbis-java (libvorbis-1.1.2) to
compress in OGG files, javaFlacEncoder 0.3.1 for the FLAC algorithm
and Java Speex Encoder v0.9.7 indicated as SPX in the figure. Then,
Alice could explore which one is more energy efficient for her
application. For instance, if it is a media store for managing music
songs, the typical file sizes could be between 15-35Mb, so she only
would need to look at this information. As can be observed in the
graphic for these sizes, the most energy efficient algorithm is LAME
to compress in MP3 files since it consumes less than 0.3 Watts,
meanwhile the other three consume more than 0.6, the double.

However, the analysis of this information in an isolated way is not enough to let
Alice takes a decision of which algorithm to use for her application
since the compression concern will not be used alone. Typically, the
Media Store will compress the audio files before uploading them
to the server. Then, in order for Alice to reason in a proper way,
she needs to know the total power consumption to both compress the file
and to send it to the server. Notice that the different compression
algorithms produce compressed file of different sizes, and therefore
the energy consumption for the communication concern will be different depending on the compression algorithm previously used.
Then, in order to simulate the energy consumption of both concerns
working together we use the Palladio Power Consumption Analyzer.

\subsection{Simulation with the Palladio Analyzer}
As previously described, we have in the HADAS repository the architectural
components modelling the energy consuming concerns
with all their respective alternatives, and with information about how they are connected
between them. These models are defined in the Palladio PCM repository diagram. Then,
for every component we have also defined its behavior in a Service Effect
Specifications (SEFF) PCM diagram.
%MONICA: Se debería referenciar a la figura de la sección 4 que falta, no?
In these diagrams we have also included
the energy models that represent the power consumption of the
internal actions (or methods) of the behavior, which have been
previously calculated taking the measures from the experiments with Joulmeter, as
explained for the compression algorithms.

\begin{figure}
\includegraphics[width=\columnwidth]{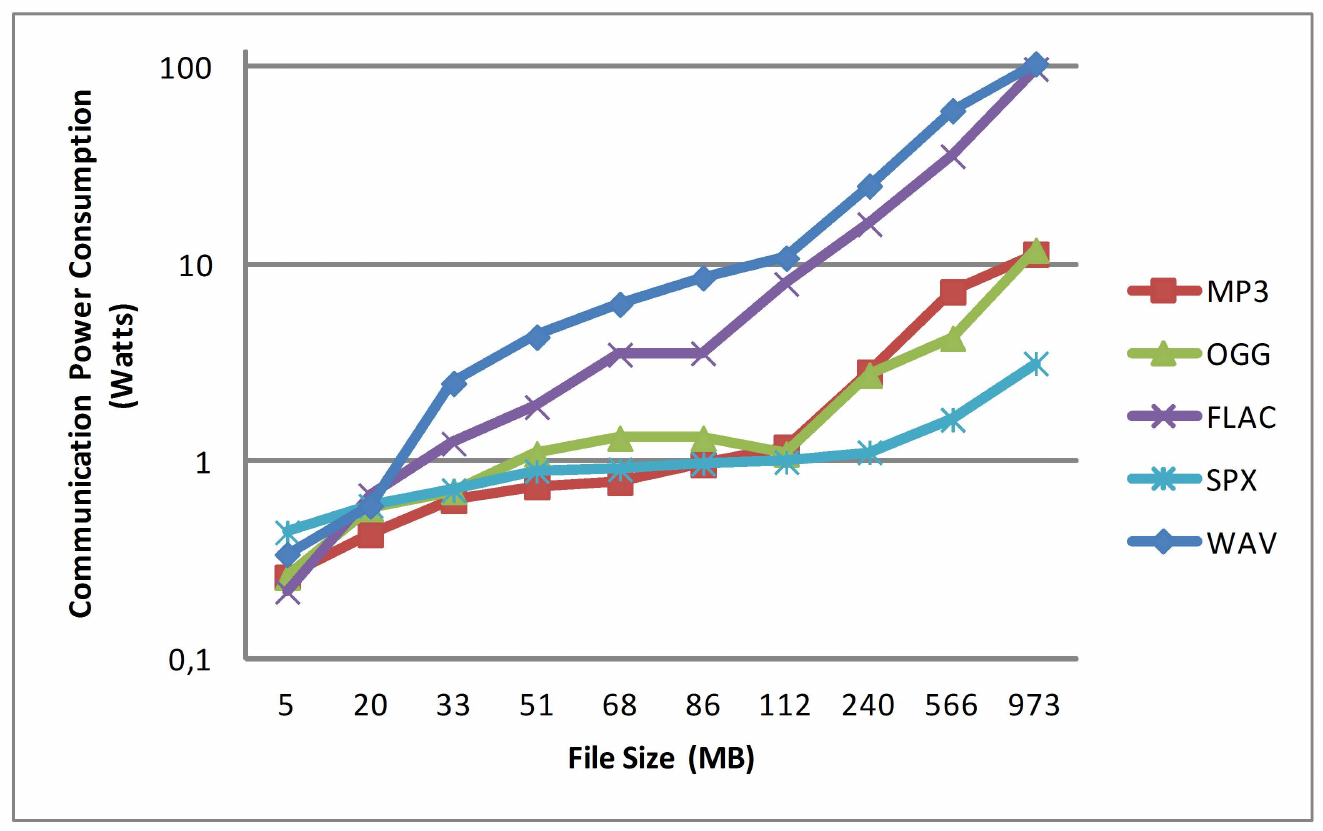}
\caption{Compression and Communication Power Consumption}
\label{fig:graphics_communication}
\end{figure}

Then, using the Palladio Power Consumption Analyzer we can simulate
how much energy the two concerns working together consume, as shown
in Figure \ref{fig:graphics_communication}. Note that this is not so
simple as sum up the Watts of the two actions (compression and
communication) since the communication energy waste depends on the
size of the file to be sent and this size strongly depends on the
compression algorithm used. In this graphic, we have also included
the power consumption of sending a WAV file without compression. We
can observe that for not too big files (less than 20Mb) it could be
more energy efficient to send the file without been compressed than
to compress it using either the javaFlacEncoder 0.3.1 for the FLAC
algorithm or the Java Speex Encoder. Then in this case, for the
typical file size of WAV song files the decision is not so clear as
before. For instance, for 30Mb files the energy consumption of first
using LAME, Vorbis and Speex to compress the file and then upload
the compressed file to the server is very similar, so Alice could
chose the one that fits better with other requirements as, for
instance, the audio quality. However, she could avoid the FLAC
algorithm for her media store application. But, if the media store
were dedicated to manage short audio recorded messages with size of
less than 10Mb this algorithm is the more energy-efficient to
compress and to send the file to the server. However, the FLAC
algorithm is the one that more power consumes for big files. Then,
for a media store to manage long audio conferences it would consume
much less the Speex algorithm. These differences of energy
consumption were not so high when we simulated the compression
algorithms alone.

Therefore, thanks to our proposal, where we use also Palladio to
make the previous simulations, Alice is able to explore all the
alternatives for all the energy-consuming concerns available in our
repository. Of course, our assistant will show her all the
alternatives for the five concerns she selected, not only for these
two. So she will be able to analyse the simulations and pick the
more proper and energy-efficient alternatives for the whole
architecture.

\subsection{Discussion}
Performig the experiments for all the individual concern and the later simulations using
Palladio for all the possible configuration of the concerns are time-consuming and not easy tasks.
However, these tasks have to be done just once when adding the concern to the repository. Then, this information can be (re)-used for many different applications, having many
advantages for the application developers:

\begin{enumerate}

\item It helps them to identify potential energy hotspots in their
application, just looking into the requirements and in the HADAS
assistant form. Then, they could be aware of them when implementing the application.
\item It detects the dependencies between
the concerns automatically, helping the developers to take into account other
hotspots that previously were not identified, as happened before to
Alice with the \textsf{Communication} concern.
\item It allows them to explore a
high number of alternatives at a glance, just making a few clicks in
the form. In \cite{Stier2015}, in order to justify the functioning
of the Palladio Power Consumption Analyzer, the authors change the
LAME algorithm for the Vorbis algorithm in their media store. But
this is done manually and the simulation of the whole architecture
has to be performed as a consequence of changing only one component
for another. With our assistant Alice has to perform just 4 clicks
to be able to explore the 4 algorithms (2 algorithms more than in
\cite{Stier2015}) and to decide which one to use for her
application. Then, the HADAS assistant provides her the PCM system
diagram with the concrete alternatives she has chosen. Without our
assistant instead of making 4 clicks, she would have to test
manually how many energy the different compression algorithms waste,
and then to modify the original media store architecture by hand to
simulate the 4 different whole architectures. Morever, if she wants
to explore the variability in other concerns the number of possible
architectures grows exponentially, which it would be very difficult
to manage.
%MONICA: No queda claro porque PCM tiene que hacerlo manualmente y nosotros generamos las arquitecturas de forma automática, si lo que estamos usando al final es PCM. Creo que esto habría que dejarlo mas claro en la sección 4. Nadia, no se si cuando añadas la arquitectura en PCM se puede decir algo concreto. Igual ya está porque explicamos que el CVL está conectado con la arquitectura base, pero se podría poner de forma más explícita.
\item Our results are accurate to decide which architectures are
more energy efficient than others. In \cite{Stier2015} authors
demonstrate that Palladio Power Consumption Analyzer is suited to
accurately predict energy consumption on an architectural level.
Therefore, as we are using this tool to make our simulations we can
conclude that the results we offer are also accurate. Anyway, as we
have explained several times, our purpose is not to predict the
exact number of watts wasted for a variant, but  which one could be
more energy-efficient than other.

\end{enumerate}

%In this section we will evaluate our HADAS Green tool using different approaches: (i) we will first analyse the degree of auto-mation that …; (ii) we will discuss the great benefits of having built-in the energy hotspots dependencies, versus identifying and introducing them manually by the software architect; (iii) we will also show the usefulness of energy graphs that compare different alternatives for each energy hotspot; (iv) xxx and, (v) finally, we will analyse the correctness of our tool results based on PCM tool
%
%\subsection{Degree of Automation}
%Generative approaches in software engineering automatically generate models or code, with the goal of minimizing the developer effort ……
%
%\subsection{Benefit of the Dependencies Exploitation}
%In addition, our tool specifies …..
%
%\subsection{Comparting Energy Consumption of Alternatives}
%One of the most interesting outputs of our tool are the energy graphs showing the …..
%
%\subsection{Simulations to Know in Advance the Energy Consumption}
%Instead of manually simulate …
%
%\subsection{Correctness of the Simulations}
%We have used PMC tool ….
%We consider that the results of the evaluation is ….

\section{Conclusions}

In this work we have presented the HADAS Green Assistant, a tool that helps application developers to be energy-aware when they are designing their applications trying to produce energy-efficient software. In order to develop the green assistant we have built a green repository composed by energy consuming concerns. These concerns represent the different ways of implementing the energy hotspots we have detected in nowadays applications, as data storage and access, communication, compression and so on. 

We have modeled the concerns using variability models to manage the different implementations. The variability models represent both all the possible alternatives for each concern and the dependencies between them. It is really necessary to take into account these dependencies because we want to offer to the application developers the possibility to reason about energy consumption of the whole software architecture, and the concerns are working together for one application, not in an isolated way. Therefore, to provide the power consumption of every concern throw experimental tests it is not enough. We have conducted also simulations in Palladio to store in the HADAS repository the energy functions associated to any possible configuration of all the concerns working together. 

The HADAS green assistant offers also a very intuitive graphical user interface based in forms that the application developers have to fill in order to select which energy consuming concerns they want to consider and which alternatives they want to explore. 

About future works, we plan to exploit the variability models representing the energy consuming concerns to used them also at runtime. It is well known that the real energy consumption of an application strongly depends on the data used and on the final usage of such application. Then, maybe the decisions taken at design time are not enough to build real energy-efficient applications. So, we propose to use variability models at runtime to be able to reconfigure dynamically the applications to adequate their concerns to new situations.

%\appendix
%\section{Appendix Title}

\acks

Research funded by the Spanish TIN2015-64841-R (co-funded by EU
with FEDER funds) and MAGIC P12-TIC1814 projects and by the International Campus of Excellence Andalucia TECH.

% We recommend abbrvnat bibliography style.

\bibliographystyle{abbrvnat}

% The bibliography should be embedded for final submission.

%\bibliographystyle{abbrv}
\bibliography{HADASbibliography}

%\begin{thebibliography}{10}
%\softraggedright

%\bibitem[Smith et~al.(2009)Smith, Jones]{smith02}
%P. Q. Smith, and X. Y. Jones. ...reference text...

%\end{thebibliography}

\end{document}